\documentclass[ammsymb,pre,aps,twocolumn,superscriptaddress,showpacs]{revtex4}

\usepackage{graphicx}

\begin{document}

\title{The underlying complex network of the Minority Game}
\author{Inés Caridi}
\email{inescaridi@yahoo.com.ar, ceva@tandar.cnea.gov.ar}
\affiliation{Departamento de F{\'{\i}}sica, Comisi{\'o}n Nacional de Energ{\'\i }a At{\'o}%
mica, Avenida del Libertador 8250, 1429 Buenos Aires, Argentina}
\author{Horacio Ceva}
\affiliation{Departamento de F{\'{\i}}sica, Comisi{\'o}n Nacional de Energ{\'\i }a At{\'o}%
mica, Avenida del Libertador 8250, 1429 Buenos Aires, Argentina}
\date{\today}

\begin{abstract}
We study the structure of the underlying network of connections in
the Minority Game. There is not an explicit interaction among
the agents, but they interact via global magnitudes of the model and
mainly through their strategies. We define a link between two agents
by quantifying the similarity among their strategies, and analyze
the structure of the resulting underlying complex networks as a function
of the number of agents in the game and the value of the agents' memory,
in games with two strategies per player. We characterize the different
phases of this system with networks with different properties, for this
link definition. Thus, the Minority Game phase characterized by the presence
of crowds can be identified with a small world network, while the
phase with the same results as a random decision game as a random
network. Finally, we use the Full Strategy Minority Game  model,
to explicitly calculate some properties of its networks,
such as the degree distribution, for the same link definition, 
and to estimate, from them, the properties of the networks
of the Minority Game, obtaining a very good agreement with its measured properties.
\end{abstract}

\pacs{02.50.Le, 89.65.Gh, 89.75.Fb, 89.75.Hc}
\maketitle


\section{Introduction\label{seccion_Introduction}}

The Minority Game (MG) was introduced in 1997 by Challet and Zhang
\cite{challet-zhang-1} as an attempt for catching some essential
characteristics of a competitive population, in which an individual
achieves the best result when he manages to be in the minority group.
The Minority Game results itself an example of a complex system, adaptive,
agents based, with emergent properties like coordination among the
population.
In the MG there are $N$ agents that in each step $t$ of the
game must choose one of two alternatives ($0$ or $1$, for example).
Let $N_{0}(t)$ and $N_{1}(t)$ be the number of agents choosing alternatives
$0$ and $1$, respectively, such that $N_{1}(t)+N_{0}(t)=N$ for all values 
of $t$. The winers are
those that happen to be in the minority group for this time step $t$.

The history of the game is a sequence of $m$ bits with the last $m$
sides that turned out to be the minority side ($m$ is identified
as the memory of the agents). Agents play using strategies; a strategy
is a function that assigns a prediction for each one of the possible
histories. In this way, in a game with memory $m$, there are $2^{m}$
different histories, and $\mathcal{L}=2^{2^{m}}$possible strategies.
The set of $\mathcal{L}$ strategies define the so called Full Strategy
Space (FSS). Each agent has $s$ strategies, randomly chosen at the
beginning of the game from the FSS (with reposition). The game is
adaptive because in every step agents choose, among their $s$ strategies,
the one that has predicted more times the minority side. There is
a system of rewards that in each time step gives a point to the winner 
agents; moreover, there also is a system whereby in every time
step one \emph{virtual point}  is assigned to all the strategies that
correctly predict the minority side, regardless if the strategy was used
or not in this step.

The parameters of the model are $N$, $m$ and $s$. The more studied
variable of the MG is 
$\sigma=\sqrt{\frac{1}{T}\sum_{t=1}^{T}(N_{1}(t)-N_{0}(t))^{2}}$,
a measure of the cooperation of the agents, in the sense of achieving
a better use of resources from the population \cite{challet-zhang-2}.
When $|N_{1}(t)-N_{0}(t)|$ is minimum, the minority group is the greatest
possible, then points assigned to the population are as great as possible,
and the contribution
to $\sigma$, minimum. For each one of the $T$ time steps of one
realization of the MG, we obtain a data set of the variable $N_{1}(t)-N_{0}(t)$.
Then, $\sigma^{2}$ is an estimator of the second moment of the probability
distribution of the variable $N_{1}(t)-N_{0}(t)$ \cite{sigma}. For a game with
$s$ fixed, in principle, $\sigma$ would depend of two parameters,
$N$ and $m$, but it was found that $\mu_{r}=\sigma^2/N$ shows scaling
as a function of $z=2^{m}/N$ \cite{manuca-ptd}. There are three regions
in the behavior of the model: \emph{(i)} a phase in which $\mu_{r}>\mu_{r}^{rg}$
(where $\mu_{r}^{rg}$ corresponds to a game where decisions from
agents are taken at random). This is the regime where the players
performance is the worst, because overall, the population receives
fewer points. The dynamics that characterizes this phase is such that
in some steps players go in crowds to one of the sides. \emph{(ii)}
The second phase, where $\mu_{r}<\mu_{r}^{rg}$, is the region where
the population achieve more points, although the spread in the distribution
of points of the agents is the greatest. \emph{(iii)} Finally, a region
where $\mu_{r}\simeq\mu_{r}^{rg}$ in which, the rules of the MG gives
the same result as in a game where the decisions are taken at random.

The MG has been studied using different tools: numerical simulations
\cite{challet-zhang-1}-\cite{manuca-ptd};
a generalization with a temperature like variable \cite{MG-temperatura}:
mapping of the model to a spin glass \cite{MG-spin-glass}, a generalization
that includes explicit interactions between agents and exchange of
information \cite{moelbert-rios}-\cite{caridi-2};
mapping to an ensemble statistical model, the Full Strategy Minority
Game (FSMG) \cite{caridi-1}, and others.

In recent years, from many areas of science, there has been much interest
on the properties of complex networks of very different complex systems 
including social, comunication, biochemical, ecosystems and internet
networks \cite{review-barabasi}-\cite{strogatz-SW}.
In this work we address two questions that are usually raised
for other complex systems: How is the underlying complex network that
connects players in this model? and, Is the structure of this network
related with different properties of the model? The first step was
to formalize a connection among the MG agents, in order to define the
underlying network. We define a link among two agents by quantifying
the similarity between their strategies; in this form, we define a
static link (because it will not change along the game), which characterizes
the game for the parameters $N$ and $m$. Then we studied different
properties of the resulting networks. In Section 2 we will define the
link, and the properties studied. In Section 3, we will analyze the
network of the FSMG, given the same definition of the link we use
for the MG, and we analytically calculate the degree distribution
of these networks as a function of $m$. Finally, we will describe
how we estimated the degree distribution of the MG networks, from
the known results of the FSMG. In Section 4 we present our conclusions.

\section{The underlying network of the MG\label{seccion_red_MG}}

Different parameters of the model ($N$, $m$ and $s$) lead to games
with players more or less \emph{connected}. We chose to work with
the static network, \emph{i.e.} the underlying network that relates
the players since the beginning of the game, when they receive their
randomly chosen strategies.

Before introducing the definition of link used, we need to define
the Hamming distance between two strategies. The Hamming distance $d$
between a pair of strategies is the number of bits in which they differ,
normalized by the length of the strategy (measured by the number
of bits). In this work we will use the following definition of the
link:

\emph{there is a link between a pair of agents $i$ and $j$ if
between each pair of strategies (taking one of each agent) there is
a Hamming distance less than $d=1/2$.}

Given a set of nodes of the network (the $N$ players of the game)
and a set of $K$ links or connections, we define an associated network
of the MG, $G^{MG}(N,K(N,m,s))$, that depends both on the definition
of link used, and on the initial conditions (\emph{i.e.} the random
assignation of strategies to the agents). Our definition of link leads
to an unweighted, undirected network. We studied the properties of
$G^{MG}$ as a function of $N$ and $m$, and $s=2$ always. To find
the network of connections, we generate an allocation of strategies
for both values $N$ and $m$ and look at the resulting connections.
We studied these networks for different values of the parameters $m$ 
(between 2 and 14), and $N$ (101, 501 and 1001). We averaged results from 
different initial conditions.

In the following subsections, we study various properties of these
networks, including the degree's distribution, the clustering coefficient,
and the average minimum path.

\subsection{Degree distribution\label{subseccion_dist_grados} }

\begin{figure}
\includegraphics [width=10cm]{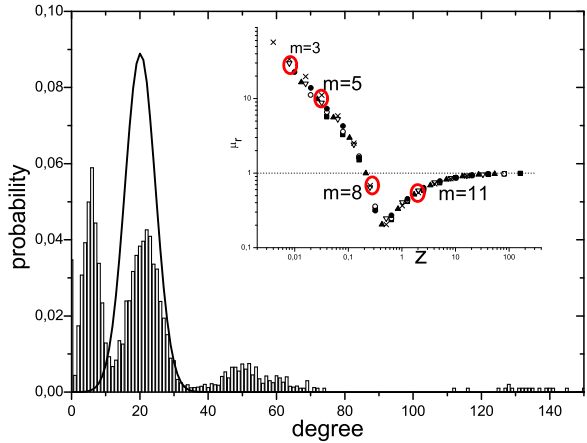}
\caption{Degree distribution histogram of MG networks for $m=3$ and $N=1001$,
averaged over 5 different initial conditions. In the inset we show the location 
of the cases illustrated in Figs.\ref{sh_high_h}-\ref{N_mil_m_once}, in the
well known diagram showing the scaling of $\mu_r$ with $z$ (scaling is shown
for   101 $\leq N \leq$ 1001, and $m \leq 14$).
Note that the probability of finding nodes of degree equal to $0$
is approximately 0.035}
\label{sh_high_h}
\end{figure}

\begin{figure}
\includegraphics[width=9.5 cm]{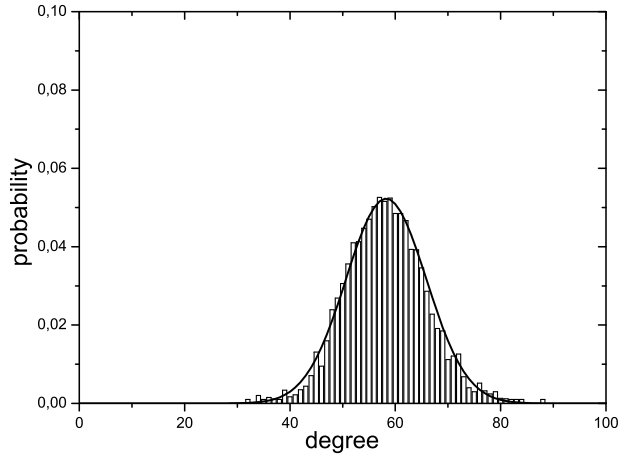}
\caption{Degree distribution Histogram of MG networks for $m=11$ and $N=1001$.}
\label{N_mil_m_once}
\end{figure}

We study the distribution of degrees (\emph{i.e.} the distribution of 
the number of links of the nodes of $G^{MG}$), for different values
of $N$ and $m$, and compare them with the known
distribution of degrees of the random network of Erdos-Renyi, $G^{ER}(N,K)$,
obtained for the same values of $N$ and $K$, but with links randomly
allocated between pairs of nodes. The link definition presented here
leads to a set of connections that grows with $m$. For
small values of $m$, the degree distributions are very different
from that of a random network, whereas as $m$ grows, they look like
those of a random network. In Figs.~\ref{sh_high_h}-\ref{N_mil_m_once} 
we present four cases for different
values of $m$ ($m=3$, $5$, $8$ and $11$) and $N=1001$; the continuous
curve is the theoretical degree distribution for an Erdos-Renyi network.
The case of $m=3$ shows five peaks; this strange degree distribution
can be explained using the mapping to the FSMG model (see section
\ref{seccion_FSMG}).

We calculated the chi square statistical hypothesis test, to learn
if the degree distribution obtained for $G^{MG}$ is consistent with
that of the $G^{ER}$. The significance values found allows us to
tell that the hypotesis is rejected for $m<11$, but it pass the test
for $m=11$.

\subsection{Clustering and minimum mean path\label{subseccion_clustering}}

We also calculate the clustering coefficient and the minimum average
path for $G^{MG}$. Given a node $\nu$ with $k_{\nu}$ links with the other nodes,
the clustering coefficient meassures the fraction of edges that actually 
exist between these nodes, out of all the possible edges that could exist
between them. The global clustering $C$ is an average of the clustering 
coefficients calculated by considering those agents that have a degree 
greater than one.

\begin{figure}
\includegraphics[width=9.5cm]{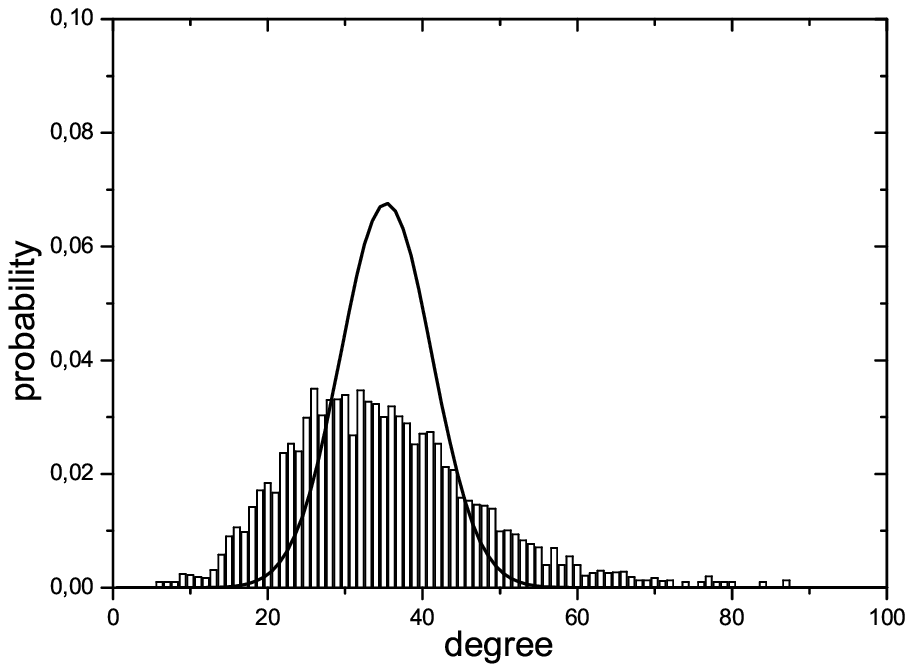}
\caption{Degree distribution Histogram of MG networks for $m=5$ and $N=1001$}
\end{figure}

\begin{figure}
\includegraphics[width=9.5 cm]{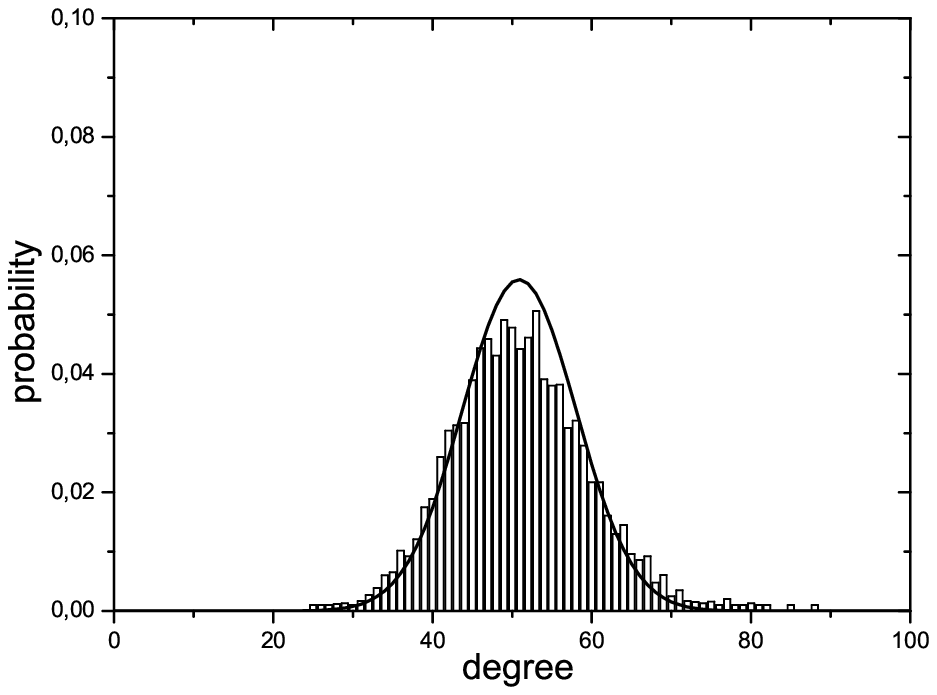}
\caption{Degree distribution Histogram of MG networks for $m=8$ and $N=1001$.}
\end{figure}

\begin{figure}  
\includegraphics[width=9.5 cm]{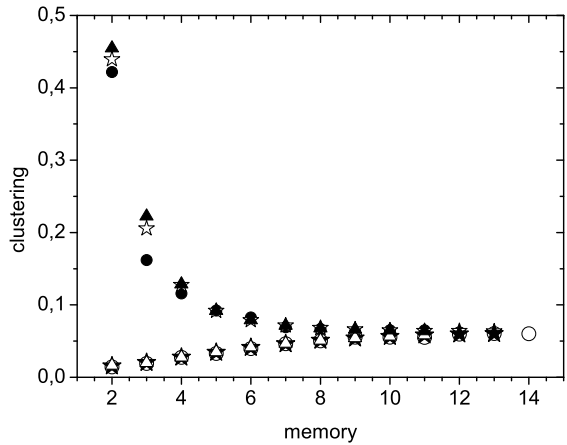}

\caption{Clustering coefficient as a function of $m$ for games with different
number of agents. In all cases the results shown are averages over 10 different 
initial conditions. MG data: Filled circles, empty stars and filled triangles
for $N$=101, 501 and 1001 respectively. Random games data: empty circles, filled stars
and empty triangles for $N$=101, 501 and 1001 respectively.}
\label{fig_clustering}
\end{figure}

We study $C$ in terms of $m$, for several values of $N$,
and compare our results with the known values for $G^{ER}$. We found that
$C$ is a function of $m$, but does not depends on $N$. 
We found that the MG clustering is much greater than the
corresponding value for random networks for values of memory up to
$m=8$; for bigger values of $m$, both coefficients are similar,
as shown in Fig.\ref{fig_clustering}. As observed in real networks,
high clusterings are typical of networks where the link represents
a social relationship, as the networks of friendships (it is very
likely that somebody's friends are also friends between themselves). In
the case of MG, we see this feature for small values of $m$, where
there are crowds effects that make the system inefficient
in the use of resources. As $m$ grows, although the number of connections
increases, clustering reflects that these connections are allocated
without transitivity, \emph{i.e.} the probability that two neighbors
of a node are connected to each other, is the same as the corresponding
probability for a random network.

\begin{figure} 
\includegraphics[width=9.5cm]{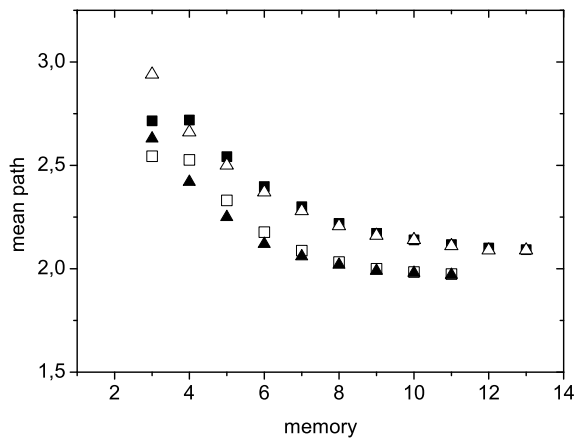}
\caption{Minimum mean path as a function of $m$ for games with different
number of agents. MG data: filled and empty squares, for $N$=501 and 1001, respectively;
Random games data: empty and filled triangles, for $N$=501 and 1001, respectively.  
As before, the data shown is an average over 10 different initial conditions.}
\label{fig_caminos}
\end{figure}

We calculate the average shortest path length between two nodes in
 $G^{MG}$ and compare
it with the corresponding values for $G^{ER}$, for all values of
$N$ and $m$ we work with. In Fig.\ref{fig_caminos} we can
see that the minimum path of $G^{MG}$ coincides with that of $G^{ER}$
for almost all values of $m$; moreover notice that this magnitude 
depends on both $N$ and $m$.
Given both properties, clustering and minimum average path, we can
say that for small memory values the MG network is a small world, while
for greater values of $m$, $G^{MG}$ behaves as a random network.%

\section{Network of the Full Strategy Minority Game model\label{seccion_FSMG}}

The FSMG is a model introduced in \cite{caridi-1} as a mapping of the
MG to other model with the same rules, but where its $\mathcal{N}$
agents are chosen in a such a way that all possible
different agents that could exist, choosing strategies from the FSS, are present. 
Thus, for a game with $s=2$ strategies for player,
there are as many agents as the number of all possible pairs of strategies
that can be formed (with reposition) from the FSS of $\mathcal{L}$
strategies. Therefore, the number of players will be:

\begin{equation}
\mathcal{N}(m)=\left(\begin{array}{c}
\mathcal{L}\\
2\end{array}\right)+\mathcal{L}=\left(\begin{array}{c}
2^{2^m}\\
2\end{array}\right)+2^{2^m}\label{agentes virtuales}\end{equation}

For simplicity, we write $\mathcal{N\equiv N}(m)$ in the following.

In a game with $s=3$ strategies for agent, the number of players
of FSMG equals the number of groups of three strategies from $\mathcal{L}$,
and so forth. In this case, we work with $s=2$.

We studied the properties of the complex network of FSMG, given the
same link definition between agents that has been used to assemble
networks of MG. By definition, this network, $G^{FSMG}$, has $\mathcal{N}$
nodes. As $\mathcal{N}$ is only a function of $m$, the set of connections,
$K$, is also a function only of $m$. Then, the network and its properties
only depend on $m$, \emph{i.e.}: $G^{FSMG}($$\mathcal{N}(m),K(m))$. For
example, in a game with $m=2$, FSMG has $\mathcal{N}=136$ agents (then $G^{FSMG}$ 
will have $\mathcal{N}=136$ nodes). In the following subsection,
we describe the analytic calculation of the degree distribution
of $G^{FSMG}$ as a function of $m$.

\subsection{Analytic calculation of the degree distribution of the FSMG model
\label{subseccion_calculo_FSMG}}

By definition, the FSMG model contains all possible agents that can
be formed from the complete set of $\mathcal{L}$ strategies (each
agent has $s=2$ strategies from $\mathcal{L}$, with reposition).
As it happened before (see reference \cite{caridi-1}), we again
benefit from the symmetry associated with this model, in which all
possible strategies are present. As a result of this symmetry, the
degrees associated with the FSMG network can just take only a few
values. In fact, the amount of different values is, \emph{at most},
$2^{m}+1$. For $m=2$, for example, there only exist nodes with degree
0, 2, 3 and 14 in the FSMG network.

We define $\mathcal{N}(m,d)$ as the number of nodes whose pairs of strategies
have among them a Hamming
distance equal to $d$. The expressions for $\mathcal{N}(m,d)$ are

\begin{equation}
\mathcal{N}(m,d)=2^{(2^{m}-1)}\left(\begin{array}{c}
2^{m}\\
2^{m}d\end{array}\right)\:\: for\:\: d\neq0\label{N(m,d)-FSMG}\end{equation}

\[
\mathcal{N}(m,d)=2^{2^{m}}\:\: for\:\: d=0.\]

such that

\begin{equation}
\sum_{d\;\in\;\left\{ 0,\frac{1}{2^{m}},...,1\right\} }\mathcal{N}(m,d)=
\mathcal{N}(m)\label{normalizacion_N}\end{equation}

In Ec. \ref{normalizacion_N}, the sum is over the possible values
that can take the variable $d$ ($0,\frac{1}{2^{m}},...,1)$. It will
be usefull to define $p(m,d)$, as

\begin{equation}
p(m,d)=\frac{\mathcal{N}(m,d)}{\mathcal{N}}\label{p(m,d)}\end{equation}

$p(m,d)$ represents the probability of finding a node in the FSMG
network, for a game with memory $m$, whose strategies have a Hamming
distance given by $d$.

As we will see, all nodes belonging to the same subset $\mathcal{N}(m,d)$
will have the same degree, that we will call $k(m,d)$. Hence, the
maximun number of possible degree values is the number of possible
values that can take the variable $d$, namely $2^{m}+1$ values, \emph{i.e.}
the different subsets of nodes that could exist.

Given a node of the subset $\mathcal{N}(m,d)$, there will be a subset
of strategies from which we can choose to form nodes that have a connection
to this node. We define the size of this strategies subset as $E(m,d)$,
that only depends of the values of $m$ and $d$.
Taking two strategies of this subset, with reposition, we obtain a
node that will have a link with each of the nodes of the subset $\mathcal{N}(m,d)$.
Hence, all the nodes belonging to the subset $\mathcal{N}(m,d)$ will have the same degree. 
 Knowing $E(m,d)$, we can calculate the degree
of the nodes of the subset $\mathcal{N}(m,d)$, as the number of pairs
that can be formed (with reposition) from the subset $E(m,d)$, which
will be:

\begin{equation}
k(m,d)=\frac{E(m,d)*(E(m,d)-1)}{2}+E(m,d)\label{k(m,d)}\end{equation}

The only consideration is that, sometimes, when the nodes of the subset
$\mathcal{N}(m,d)$ can be linked with nodes of the same subset, we
must discount a node (a couple of strategies) in $k(m,d)$, in order
not to consider the link between a node with itself.

The calculation of $E(m,d)$ can be found in the Appendix.

In the inset of Fig.\ref{pico}, we show the
actual degree distribution for $G^{FSMG}$, for $m=3$, called $P(k)$.
Even though the maximum number of possible peaks can be $2^3+1=9$, in this 
case $P(k)$ has only five peaks. This reduction is due to the fact that subsets
$\mathcal{N}(m,d)$ for differents values of $d$ can have $E(m,d)=E(m,d')$ and, 
therefore, the same degree.

\subsection{Estimating parameters of the networks $G^{MG}$ from those of the
networks $G^{FSMG}$\label{subseccion_estimacion} }

From the analytic calculation of the degree distribution of $G^{FSMG}$,
we can estimate the degree distribution of $G^{MG}$, for a given
value of $N$. The set of nodes of $G^{MG}$ is a sample of all nodes
of the FSMG network. This sample is selected at random, because the
assignation of strategies to agents of the MG is done randomly. Thus,
one has to choose $N$ nodes at random, with replacement (because
in the MG it can be identical players) from the set of $\mathcal{N}$
nodes of the $G^{FSMG}$.

By choosing only $N$ nodes, we are considering  a subset of nodes and
a subset of connections, \emph{i.e.} a subgraph induced by the set
of nodes $N$. The relationship between $N$ and $\mathcal{N}$ tells
 how representative is this sample. We call $q=N/\mathcal{N}$
the probability that a node of the FSMG be elected to form the network
of MG. The question we are trying to answer is how to infer the degree
distribution of $G^{MG}$ (that we call $P(\widetilde{k})$) from $P(k)$,
the known degree distribution of $G^{FSMG}$.

As mentioned before, the FSMG networks have a discrete degree distribution,
 $P(k)$, that can take at most $2^{m}+1$; in fact not only the values, but also 
 their location and height depends only of $m$.
The degree distribution of MG network will also be discrete (by definition
of degree), but the variable may take any integer value, $\tilde{k}$
, with certain probability $P(\widetilde{k})$. Going from FSMG to
MG, each peak of $P(k)$ becomes a probability distribution.

We will first analize how one peak of $P(k)$ becomes a probability
distribution. Let consider, for instance, the peak of degree $k_{i}$.
For the subset of nodes of FSMG with a given degree value $k_{i}$,
the probability of one node being chosen to be part of the set of
$N$ agents of MG, is $q$. The probability that each of the neighboring
nodes of this node is chosen is also $q$. Therefore, the probability
distribution of degrees $\widetilde{k}$ for this subset of chosen
nodes for MG that in the FSMG network had degree $k_{i}$ is now a
binomial distribution:

\begin{equation}
\left(\begin{array}{c}
k_{i}\\
\tilde{k}\end{array}\right)q^{\tilde{k}}(1-q)^{k_{i}-\tilde{k}}
\label{distribucion_picos}\end{equation}

whose mean value and variance are:

\begin{equation}
<\widetilde{k}>=k_{i}q\label{estimacion de p}\end{equation}

\begin{equation}
(\sigma_{\widetilde{k}})^2=k_{i}q(1-q)\label{estimacion del sigma}\end{equation}

We use the Central Limit theorem to aproximate the binomial distribution
by a gaussian distribution, with the same values of variance and
mean:

\begin{equation}
\frac{1}{\sqrt{2\pi}\sigma_{\widetilde{k}}}
e^{-(\widetilde{k}-<\widetilde{k}>)^{2}/2\sigma_{\tilde{k}}^{2}}
\label{pico_gauss}\end{equation}

In our calculations for $m > 2$ it is  $q<<1$, so we approximate $\sigma_{\widetilde{k}}\simeq\sqrt{k_iq}$.

Finally, all peaks of $P(k)$ will contribute to $P(\widetilde{k})$,
which results:

\begin{equation}
P(\widetilde{k})=\sum\limits_{i}
P(k_{i}) \frac{1}{\sqrt{2\pi k_{i}q}}
e^{-(\widetilde{k}-k_iq)^{2}/2k_{i}q}  
  \label{P_ktilde}\end{equation}
  
where the sum is over all the peaks of P(k).

\begin{figure}
\includegraphics [width=8.5cm]{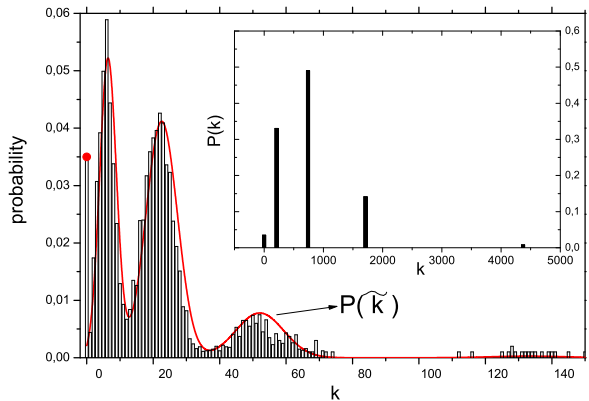}
\caption{Degree distribution for MG networks, for $N=1001$ and $m=3$. 
The histogram is an average over 10 different realizations. The continuous 
line shows the distribution estimated using the FSMG, see Eq. \ref{P_ktilde}. }
\label{pico}
\end{figure}

We compare the MG network degree distribution estimated from the
FSMG, with that obtained from realizations of networks of the MG,
in a game of $N=1001$ agents. In Fig.\ref{pico}
it is possible to see the very good adjustment found for 
$N=1001$ and $m=3$. It is remarkable, in particular, the excelent 
fit obtained for the probability to have a  
disconected node in the MG network, $P(\widetilde{k}=0)=P(k=0)=0.035$

\section{Conclusions\label{seccion_Conclusions}}

As far as we know, this work is the first attempt to caracterize the implicit
interactions between MG agents, as a complex network.

We formalize an underlying network for the Minority Game, by quantifying
the similarity of the strategies between two agents. Given the resulting
definition of the link, one can see that in the MG phase characterized
by the presence of crowds,  its network can be identified
as a small world, whereas in the phase where the system behaves
like in a game of random decisions, the underlying network behaves as a 
random one, with the same clustering coefficient, degree distribution
and minimal path. We analytically calculate the degree distribution
for the underlying network of the FSMG model, and from this 
distribution, we estimated the degree distribution of MG networks,
with a very good agreement. This reflects, again, that the FSMG is
a useful model to study the MG.

It would be interesting to explore in the future the effect of employing weigthed
links in these networks. And in those cases where an explicit interaction 
between some agents are introduced, it could be of some help to
add together the effects of the networks of explicit and implicit interactions.

\section{Appendix\label{seccion_Appendix}}

In the following we will explain in detail the form in which we did
the calculation of $E(m,d)$. As mentioned in \ref{subseccion_calculo_FSMG},
$E(m,d)$ is the size of the set of strategies from which one can
choose to form nodes that have a link with some node of the subset
$\mathcal{N}(m,d)$.

We defined a link between two agents when, among each pair of strategies
(one of each agent), there is a Hamming distance $d<1/2$. Let us call $\delta$
the threshold, such that the previous condition became $d\leq\delta$.
For example, for the case of $m=2$, $\delta=1/4$ and for $m=3$,
$\delta=3/8$, and in general,

\begin{equation}
\delta=\frac{1}{2}-\frac{1}{2^{m}}\label{definicion_delta}\end{equation}

To find $E(m,d)$ we consider both strategies of a node of $\mathcal{N}(m,d)$,
and count how many strategies have a normalized Hamming distance with
both strategies smaller or equal than $\delta$.

As an example, let us count the case of nodes with zero degree. These
are the nodes belonging to the subset $\mathcal{N}(m,d)$ such
that $E(m,d)=0$ and therefore do not have
connectivity. The cases of $m$ and $d$ for which $E(m,d)=0$ are
such that $d-\delta>\delta$, \emph{i.e.} $d>2\delta$. This is so
 since if the distance between two strategies is greather
than $2\delta$, it does not exist any strategy whose Hamming distance
to the couple of strategies is simultaneously smaller or equal than
$\delta$.

Therefore, for the case $m=2$:

\[
E(m=2,d=1)=0\]

\[
E(m=2,d=3/4)=0\]

Whereas for $m=3$:

\[
E(m=3,d=1)=0\]

\[
E(m=3,d=7/8)=0\]

And, for general values of $m$:

\[
E(m,d>2\delta)=0\]

\[
k(m,d>2\delta)=0\]

As another example, we will calculate $E(m,d=0)$, that is to say,
we want to find the connectivity of the nodes of the subset $\mathcal{N}(m,d=0)$
(subset of nodes formed by two equal strategies). In this case one
has to count all the strategies that differ from the node's strategies
in a bit, two bits, three bits ... up to $2^{m}\delta$ bit. It is:

\[
E(m,d=0)=\sum_{i=0}^{2^{m}\delta}\left(\begin{array}{c}
2^{m}\\
i\end{array}\right)\]

Since this expression includes the strategies that forms
the node, then in the calculation of the degree of nodes of the subset
$\mathcal{N}(m,d=0)$, we discount a couple of strategies, since we
do not consider the connection of a node with itself. Therefore:

\[
k(m,d=0)=\frac{E(m,d=0)(E(m,d=0)-1)}{2}+E(m,d=0)-1\]

The case $d=2\delta$ gives:

\[
E(m,d=2\delta)=\left(\begin{array}{c}
2\delta\:2^{m}\\
\delta\:2^{m}\end{array}\right)=\left(\begin{array}{c}
d\:2^{m}\\
d\:2^{m-1}\end{array}\right)\]

When $\delta<d<2\delta$ one finds

\[
E(m,d)=\sum_{i=(d-\delta)2^{m}}^{2^{m}\delta}\left(\begin{array}{c}
d\:2^{m}\\
i\end{array}\right)\sum_{j=0}^{J(i)}\left(\begin{array}{c}
(1-d)\:2^{m}\\
j\end{array}\right)\]

where $J(i)=min\{2^{m}(\delta-d)+i;\delta2^{m}-i\}$

Finally, for the case $0<d<\delta$ the result is 

\[
E(m,d)=\sum_{i=0}^{2^{m}d}\left(\begin{array}{c}
d\:2^{m}\\
i\end{array}\right)\sum_{j=0}^{2^{m}\delta-J(i)}\left(\begin{array}{c}
(1-d)\:2^{m}\\
j\end{array}\right)\]

where $J(i)=\left|max\{2^{m}d-i,i\}\right|$

For the last three cases, the degree, $k(m,d)$, will be calculated
as in Ec. \ref{k(m,d)}.

I.C. would like to thank Prof. Hernán Solari for useful conversations


\begin{thebibliography}{99}
\bibitem{sigma}  Notice that upon averaging over different initial conditions 
$\sigma$ becomes the standard deviation, because in this case $<N_1 - N_0>=0$
\end{thebibliography}
\end{document}